# Strong-field extreme-ultraviolet dressing of atomic double excitation

**Authors:**

Christian Ott[1*], Lennart Aufleger[1], Thomas Ding[1], Marc Rebholz[1], Alexander Magunia[1], Maximilian Hartmann[1], Veit Stooß[1], David Wachs[1], Paul Birk[1], Gergana D Borisova[1], Kristina Meyer[1], Patrick Rupprecht[1], Carina da Costa Castanheira[1], Robert Moshammer[1], Andrew R Attar[2], Thomas Gaumnitz[3], Zhi Heng Loh[4], Stefan Düsterer[5], Rolf Treusch[5], Joachim Ullrich[6], Yuhai Jiang[7], Michael Meyer[8], Peter Lambropoulos[9], and Thomas Pfeifer[1#]

[*]christian.ott@mpi-hd.mpg.de

[#]thomas.pfeifer@mpi-hd.mpg.de

**Affiliations:**

[1]Max-Planck-Institut für Kernphysik, Saupfercheckweg 1, 69117 Heidelberg, Germany

[2]Department of Chemistry, University of California, Berkeley, California 94720, USA

[3]Laboratorium für Physikalische Chemie, ETH Zürich, Vladimir-Prelog-Weg 2, 8093 Zürich, Switzerland

[4]Division of Chemistry and Biological Chemistry, and Division of Physics and Applied Physics, School of Physical and Mathematical Science, Nanyang Technological University, Singapore 637371, Singapore

[5]Deutsches Elektronen-Synchrotron DESY, Notkestraße 85, 22607 Hamburg, Germany

[6]Physikalisch-Technische Bundesanstalt, Bundesallee 100, 38116 Braunschweig, Germany

[7]Shanghai Advanced Research Institute, Chinese Academy of Sciences, Shanghai 201210, China

[8]European XFEL GmbH, Holzkoppel 4, 22869 Schenefeld Hamburg, Germany

[9]Department of Physics, University of Crete, 71003 Heraklion, Crete, Greece

**Abstract:**

**We report on the experimental observation of strong-field dressing of an autoionizing two-electron state in helium with intense extreme-ultraviolet laser pulses from a free-electron laser. The asymmetric Fano line shape of this transition is spectrally resolved, and we observe modifications of the resonance asymmetry structure for increasing free-**

electron-laser pulse energy on the order of few tens of μJ. A quantum-mechanical calculation of the time-dependent dipole response of this autoionizing state, driven by classical extreme-ultraviolet (XUV) electric fields, reveals a direct link between strong-field-induced energy and phase shifts of the doubly excited state and the Fano line-shape asymmetry. The experimental results obtained at the Free-Electron Laser in Hamburg (FLASH) thus correspond to transient energy shifts on the order of few meV, induced by strong XUV fields. These results open up a new way of performing non-perturbative XUV nonlinear optics for the light-matter interaction of resonant electronic transitions in atoms at short wavelengths.

Quantum mechanics provides a consistent description of the structure and dynamics of atoms, the constituents of our macroscopic world. In particular, it describes how bound excited states in atoms are formed through the Coulomb interaction of the positively charged nucleus and the negatively charged electrons. With the obvious exception of the ground state, such states possess a finite lifetime, with singly excited states decaying through photon emission via the interaction with the radiation field. For two-electron excitations of neutral atoms, the Coulomb interaction between the electrons is much more effective such that at least one electron will eventually be ionized, which typically marks the leading contribution to the decay of the excited state for the case of light atoms. Thus ionization is a fundamentally important and very basic effect that accompanies the physics of multi-electron excitations in atoms [1]. An interesting situation arises if the interaction of such states with the radiation field is significantly increased which nowadays can be achieved by using extreme ultraviolet (XUV) or x-ray light sources. In addition, the properties of these radiation fields can often be well controlled, thus providing a unique toolbox for exploring the dynamics of excited states, e.g., by performing time-resolved investigations with lab-based attosecond high-order harmonic generation (HHG) sources [2,3], or facility-based femtosecond XUV/x-ray free-electron lasers (FELs) [4,5]. The latter deliver particularly high intensities for XUV/x-ray nonlinear optics [6] with ultrafast time resolution and site-specific core-level access [7], and nowadays even approach the attosecond regime [8].

The helium atom consists of two electrons bound to a nucleus, representing the ideal case of a Coulombic three-body system, which serves as a benchmark for developing a theoretical description [1,9,10] and most importantly also for controlling the dynamics of two bound electrons with strong external electric fields [11–13]. A unique fingerprint for the dynamics of two active electrons in helium are doubly excited states, whose spectroscopic signature manifests in asymmetric Fano line shapes [14], where the excited state rapidly decays through autoionization which is caused by the direct interaction between the two electrons.

Using femtosecond near-infrared (NIR) laser fields at moderate intensities, a control of the two-electron dynamics in helium has been demonstrated both in theory and experiment, enabled by probing with weak HHG attosecond XUV pulses [15–24]. On the other hand, strongly driving an autoionizing state directly with intense short-wavelength fields, as proposed theoretically in [11,12], has not been demonstrated experimentally. Very recently, this topic has received new theoretical interest [25–27], now coming experimentally within reach with state-of-the-art FEL pulses. The latter have nowadays indeed become the workhorse for the study of resonant nonlinear light-matter interaction in the XUV and x-ray spectral domain [28–35].

Here, we use intense FEL pulses to directly drive a two-electron transition in helium. This is conceptually realized with an absorption measurement in Fraunhofer-type transmission geometry (see Fig. 1a). That is, the FEL pulses (depicted in black color), which are spectrally wider than the resonance profile, are attenuated upon propagation through a moderately dense helium gas target. The narrow resonant absorption line can thus be spectrally resolved and is imprinted on the transmitted pulses (depicted in red color), which represents the directly measured quantity of this experiment. In our case, we tune a strong XUV pulse to an isolated dipole-allowed transition in helium, namely the $1s^2$ ($^1S^e$) to $2s2p$ ($^1P^o$) two-electron transition. It should be noted that such direct coupling of a two-electron transition with a single-frequency (within the pulse bandwidth) electric field is only possible due to the electron-electron interaction within the system. Related to the spectral bandwidth criteria discussed above, one enters a regime where the dressing XUV pulse duration is shorter than the lifetime of the excited state. The autoionization decay thus subsequently evolves in an essentially field-free environment, temporally after the initial excitation and dressing during the pulse. This is conceptually similar to Fano line-shape modifications upon strong-field NIR dressing as reported in Ref. [19]. In brief, a field-induced energy shift $\Delta E$ occurs within a finite time window $\Delta T$ during the interaction with a short pulse. This interaction may occur only within a fraction of the autoionization lifetime, which eventually translates into a phase shift $\Delta\phi$ of the absorption response that is encoded into a measureable change in the Fano lineshape asymmetry via $\phi = 2\cdot\arg[q - i]$. In the following, we consider this transient energy shift to be caused by the XUV field alone, acting at the same time as both excitation and dressing interaction, which is also depicted in Fig. 1b.

To illustrate the expected line-shape modifications of the Fano resonance in helium in intense XUV fields we start with a computational model. Following [11,14], the spectrally isolated 2s2p resonance, reached via a dipole-allowed single-photon transition from the ground state, is well represented by an expansion consisting of the ground state $\psi[1s^2]$, the doubly excited state $\psi[2s2p]$, and its interaction with the single-electron continuum $\psi_E[1sEp]$, whereby the weak-field resonant transition and configuration interaction is well known [36,37]. The total wavefunction is expanded as $\Psi(t) = c_g(t)\cdot\psi[1s^2] + c_e(t)\cdot\psi[2s2p] + \int dE\cdot c_E(t)\cdot\psi_E[1sEp]$, with the time-dependent complex-valued expansion coefficients $c_i(t)$. Within the standard approach of adiabatically eliminating the continuum {i.e., $\dot{c}_E(t) = 0$; see, e.g., Ref. [38]}, the problem essentially reduces to solving the time-dependent Schrödinger equation of a two-level system, which consists of the two bound-state two-electron configurations with coefficients $c_g(t)$ and $c_e(t)$, respectively. The system is driven by an external electric field $F(t)$ within the rotating-wave approximation, and eventually yields the complex-valued dipole response $d(t) \propto <\Psi(t)|\hat{r}|\Psi(t)>$. The absorption spectrum $A(\omega)$ is then obtained through $A(\omega) \propto \omega \cdot \acute{A}\{\tilde{d}(\omega)/\tilde{F}(\omega)\}$, (see, e.g., Ref. [39]), with the positive-frequency Fourier transform of the dipole response $\tilde{d}(\omega)$ and the electric field $\tilde{F}(\omega)$, respectively, in complex representation.

Using this model, we now demonstrate how the excitation and dressing with short XUV fields leads to a transient energy shift, which eventually translates into a modification of the Fano line shape. Therefore we calculate the time-dependent dipole response $d(t)$ of the system driven by a single Gaussian-shaped XUV pulse of 5-fs full-width-at-half-maximum (FWHM) duration, and centered close to resonance, slightly red detuned, at 60.10 eV photon energy. The interaction with the external XUV field is considerably shorter than the 17 fs exponential (1/e) autoionization decay time reported in the literature [36,37]. With the small detuning we expect a positive transient energy shift of the upper state of the coupled two-level system. The calculated XUV-intensity-dependent absorption spectrum is plotted in Fig. 2a together with selected lineouts at low, intermediate and high XUV intensity (see Fig. 2b–d). One clearly observes a decrease of the on-resonance absorbance with increasing intensity. Furthermore, the resonance shape appears more symmetric for increasing XUV intensity. This is further quantified by fitting the absorption response with the Fano line profile $\sigma(\varepsilon) = a \cdot [(q + \varepsilon)^2 / (1 + \varepsilon^2) - 1] + b$, with the amplitude $a$, offset $b$, Fano asymmetry parameter $q$ and the reduced energy $\varepsilon = 2\cdot(\omega - \omega_r) / \Gamma$ with spectral energy $E = \hbar\omega$, resonance energy $E_r = \hbar\omega_r$ and autoionization decay energy width $\hbar\Gamma$. The result of the fit, together with the $q$ asymmetry parameter, is shown in Fig. 2b–d. This quantifies and confirms the observed trend of more symmetric line shapes (i.e., increasing magnitude of $q$) with increasing XUV intensity. Now applying the conversion $\phi = 2 \cdot \arg[q - i]$ as introduced above and in Ref. [19], one can relate these asymmetry changes to phase shifts $\Delta\phi = \phi - \phi_0$, where $\phi_0 = -5.590$ rad is the offset phase of the original Fano asymmetry $q_0 = -2.77$ when the system is driven by a weak field. The resulting $\Delta\phi$'s are also reported in Fig. 2b–d. To prove that these phase shifts indeed relate to the integrated energy shift $\Delta E(t)$ upon field-dressing during the excitation, we plot in Fig. 2e the transient phase evolution $\phi_e(t) = \arg[c_e(t)]$ of the complex expansion coefficient of the 2s2p excited state. Hereby, the trivial time evolution of $\phi_e(t)$ when driven by a weak field has been subtracted, which directly reveals the dressing-field-induced phase shift shown in Fig. 2e. The numerical values, quantified temporally after the dressing, are shown and agree well with those obtained through fitting the Fano line shapes (cf. the values printed in Fig. 2b–d), revealing a common systematic increase in magnitude with the dressing field strength. The slight discrepancy on the order of a few percent can be attributed to the effect of a finite pulse duration, which is not captured within a single $q$ parameter fit of the line shape. Indeed it is well known that other parameters of the external field (pulse duration, intensity, detuning, etc.) influence the underlying strong-field autoionization dynamics [11]. We also compute $\Delta E(t) = -\hbar \cdot \partial/\partial t [\phi_e(t)]$ which directly reveals the transient energy shift during the pulse, and is plotted in Fig. 2f. As expected, we observe a shift to positive energies due to the slightly red-detuned central photon energy, where the energy shift effectively corresponds to the generalized Rabi frequency. We would like to stress that such transient energy shift on a time scale shorter than the state's lifetime is not captured by a spectral shift of the absorption line, but is directly encoded into its asymmetry profile, in close conceptual similarity to previous strong-field-NIR dressing effects reported in Ref. [19].

Now we present an experimental proof of this concept, observing Fano asymmetry changes in helium absorption spectra with intense FEL pulses. The experiment has been carried out at the focused open-port Beamline BL2 at the Free-Electron Laser in Hamburg (FLASH) [40]. The FEL is operated in single-bunch mode at 10 Hz repetition rate, and the integrated energy contained in each pulse is measured with a parasitic gas monitor detector (GMD) [41]. This detector is installed upstream of the experiment and yields pulse energies up to ~100 µJ, whereas the averaged pulse energy over all 12,500 pulses contained in the measurement is around ~75 µJ, with typical pulse-to-pulse fluctuations due to the inherent FEL generation process via self-amplified spontaneous emission (SASE). The FEL beamline transmission is estimated to be ~50 % photon flux, which is due to several carbon-coated grazing-incidence XUV mirrors for the optical beam transport. This yields ~50 µJ maximum pulse energy available on target. The central photon energy is measured to be 60.1 eV with 0.4 eV FWHM spectral intensity bandwidth of the averaged photon spectral distribution. Based on the average pulse energy and typical FEL operation conditions [42], the pulse duration, on average, is estimated at 75 fs FWHM. It should be noted however that the stochastic substructure of the FEL pulses, with spiky structures in the time domain with durations of typically only a few femtoseconds [43,44], is important and has to be considered to draw a connection to our model simulation introduced above, which will be further discussed below. With an ellipsoidal focusing mirror, a spot size of typically 25 µm FWHM is reached. The on-target photon fluence thus reaches the order ~10 J/cm², which compares well with the parameter range of the model simulation. With an attenuating gas absorber filled with molecular nitrogen, and installed upstream of the experiment, the photon flux can be continuously lowered all the way to almost zero. Using a parasitic spectrometer with a variable-line-space (VLS) grating [45], reference spectral photon distributions are recorded for each individual FEL pulse. The XUV light is transmitted through a dense helium target, contained in a gas cell at ~100 mbar backing pressure, and 3 mm interaction length, which is much smaller than the centimeter-scale Rayleigh length of our focusing geometry. With an on-resonance photoabsorption cross section of ~10 Mbarn [37,46] and following Beer–Lambert's attenuation law, optical densities (ODs) in between OD 1 and OD 2 are reached. The transmitted spectral intensity $S_{He}(\omega)$ is measured with a second VLS-grating-based spectrometer, fully resolving the 37 meV wide absorption line within the on average 0.4 eV wide XUV spectrum (see also Fig. 1a). The experimental absorbance $A_{exp}(\omega)$ is determined via $A_{exp}(\omega) = -\log_{10}[<S_{He}(\omega)>/<S_{ref}(\omega)>]$, quantified in dimensionless OD units, where $S_{ref}(\omega)$ denotes the incoming XUV spectral intensity distribution as measured with the upstream parasitic VLS spectrometer. The mean value $<...>$ over several 10's to few 100's individual single-shot FEL spectra are taken to average over slight spectral discrepancies between the two independent spectrometers. It is $A_{exp}(\omega)$ that can be directly compared to the absorption response $A(\omega)$ obtained in theory.

In Fig. 3a we show the experimental absorption spectrum $A_{exp}(\omega)$ as a function of the on-target pulse energy. A clear decrease of the resonant absorption is observed, in agreement with the model results shown in Fig. 2a, which confirms that the XUV photon fluence assumed in the model calculation agrees well with the experimental on-target parameters.

Furthermore, with increasing photon flux we observe the transition from a typical Fano shape into a more symmetric absorption line. This is further substantiated by the spectral lineouts, shown in Fig. 3b–d, at high, intermediate and low pulse energy, respectively, where $q$ is quantified again through a fit of the Fano resonance profile. A clear increase in magnitude of $q$ is observed with increasing photon flux. A conversion to field-dressing-induced phase shifts yields $\Delta\phi_a = -(10\pm50)$ mrad, $\Delta\phi_a = -(100\pm50)$ mrad, and $\Delta\phi_a = -(280\pm50)$ mrad, for on-target pulse energy $E_a = 2.5$ µJ, $E_a = 22$ µJ, and $E_a = 42$ µJ, respectively. The reported error hereby represents an order-of-magnitude estimate of the fit. The trend of an increasingly negative phase shift with increasing photon flux also agrees well with the model predictions presented in Fig. 2. We note that changes of the resonance profile have also been predicted by recent theoretical calculations [25–27], whereas here we identify the mechanism of dressing-induced phase shifts on a timescale shorter than the autoionization lifetime, which we believe is the main reason for twisting the Fano line into a more symmetric shape with higher XUV photon fluence.

Finally we would like to comment on the influence of the stochastic FEL pulse shapes. The pulse duration of the model simulation described above relates to the typical duration of an individual intensity spike contained within the FEL pulse [47,48]. Several such spikes typically are randomly distributed within the ~75-fs-timescale average pulse duration, which is considerably longer than the 17 fs autoionization lifetime of the 2s2p state. This rather rapid decay time, compared with the average pulse duration, thus imposes an effective gate on the influence of a successive intensity spike within the same FEL pulse. The stochastic FEL pulse can thus be regarded as an averaged sequence of individual "micro-experiments", where each intensity spike within the pulse creates an independent absorption signal. Hereby, the most intense such spike interacts with the largest fraction of helium atoms within the target volume, thus the measured absorption signal is expected to be dominated by the strongest intensity spike. Therefore, we can identify the few-femtosecond strong-field XUV dressing effects as introduced in Fig. 2 as the leading mechanism for explaining the experimentally observed Fano asymmetry changes with stochastic FEL pulses. Going beyond this basic model is obviously required, however, to explore further details that are contained within the measurement. For instance, looking more closely at Fig. 3, in addition to its asymmetry change, the line profile seems to also shift slightly to higher XUV energy with increasing pulse energy. To explain such energy shift, the system needs to be dressed also after the initial excitation, i.e., during a substantially longer timespan within the autoionization lifetime. Such line shifts can thus be attributed to more complex FEL pulse shapes. A detailed investigation of this and other effects however goes beyond the scope of this first report of the main observation of strong-field XUV-dressing effects that lead to Fano asymmetry changes.

**Acknowledgements:**

We acknowledge funding from the European Research Council (ERC) (X-MuSiC-616783). YHJ acknowledges for support of the National Natural Science Foundation of China (NSFC) (11420101003 and 11827806).

**References:**

[1]    U. Fano, Reports Prog. Phys. **46**, 97 (1983).

[2]    P. Agostini and L. F. DiMauro, Reports Prog. Phys. **67**, 813 (2004).

[3]    F. Krausz and M. Ivanov, Rev. Mod. Phys. **81**, 163 (2009).

[4]    C. Pellegrini, A. Marinelli, and S. Reiche, Rev. Mod. Phys. **88**, 015006 (2016).

[5]    E. A. Seddon, J. A. Clarke, D. J. Dunning, C. Masciovecchio, C. J. Milne, F. Parmigiani, D. Rugg, J. C. H. Spence, N. R. Thompson, K. Ueda, S. M. Vinko, J. S. Wark, and W. Wurth, Reports Prog. Phys. **80**, 115901 (2017).

[6]    G. Doumy, C. Roedig, S.-K. Son, C. I. Blaga, A. D. DiChiara, R. Santra, N. Berrah, C. Bostedt, J. D. Bozek, P. H. Bucksbaum, J. P. Cryan, L. Fang, S. Ghimire, J. M. Glownia, M. Hoener, E. P. Kanter, B. Krässig, M. Kuebel, M. Messerschmidt, G. G. Paulus, D. A. Reis, N. Rohringer, L. Young, P. Agostini, and L. F. DiMauro, Phys. Rev. Lett. **106**, 083002 (2011).

[7]    L. Young, K. Ueda, M. Gühr, P. H. Bucksbaum, M. Simon, S. Mukamel, N. Rohringer, K. C. Prince, C. Masciovecchio, M. Meyer, A. Rudenko, D. Rolles, C. Bostedt, M. Fuchs, D. A. Reis, R. Santra, H. Kapteyn, M. Murnane, H. Ibrahim, F. Légaré, M. Vrakking, M. Isinger, D. Kroon, M. Gisselbrecht, A. L'Huillier, H. J. Wörner, and S. R. Leone, J. Phys. B At. Mol. Opt. Phys. **51**, 032003 (2018).

[8]   N. Hartmann, G. Hartmann, R. Heider, M. S. Wagner, M. Ilchen, J. Buck, A. O. Lindahl, C. Benko, J. Grünert, J. Krzywinski, J. Liu, A. A. Lutman, A. Marinelli, T. Maxwell, A. A. Miahnahri, S. P. Moeller, M. Planas, J. Robinson, A. K. Kazansky, N. M. Kabachnik, J. Viefhaus, T. Feurer, R. Kienberger, R. N. Coffee, and W. Helml, Nat. Photonics **12**, 215 (2018).

[9]   C. D. Lin, Phys. Rep. **257**, 1 (1995).

[10]  G. Tanner, K. Richter, and J.-M. Rost, Rev. Mod. Phys. **72**, 497 (2000).

[11]  P. Lambropoulos and P. Zoller, Phys. Rev. A **24**, 379 (1981).

[12]  K. Rzążewski and J. H. Eberly, Phys. Rev. Lett. **47**, 408 (1981).

[13]  P. Lambropoulos, P. Maragakis, and J. Zhang, Phys. Rep. **305**, 203 (1998).

[14]  U. Fano, Phys. Rev. **124**, 1866 (1961).

[15]  C. A. Nicolaides, T. Mercouris, and Y. Komninos, J. Phys. B At. Mol. Opt. Phys. **35**, L271 (2002).

[16]  M. Wickenhauser, J. Burgdörfer, F. Krausz, and M. Drescher, Phys. Rev. Lett. **94**, 023002 (2005).

[17]  L. Argenti and E. Lindroth, Phys. Rev. Lett. **105**, 053002 (2010).

[18]  S. Gilbertson, M. Chini, X. Feng, S. Khan, Y. Wu, and Z. Chang, Phys. Rev. Lett. **105**, 263003 (2010).

[19]  C. Ott, A. Kaldun, P. Raith, K. Meyer, M. Laux, J. Evers, C. H. Keitel, C. H. Greene, and T. Pfeifer, Science (80-. ). **340**, 716 (2013).

[20]  C. Ott, A. Kaldun, L. Argenti, P. Raith, K. Meyer, M. Laux, Y. Zhang, A. Blättermann, S. Hagstotz, T. Ding, R. Heck, J. Madroñero, F. Martín, and T. Pfeifer, Nature **516**, 374 (2014).

[21] L. Argenti, Á. Jiménez-Galán, C. Marante, C. Ott, T. Pfeifer, and F. Martín, Phys. Rev. A **91**, 061403 (2015).

[22] V. Gruson, L. Barreau, Á. Jiménez-Galan, F. Risoud, J. Caillat, A. Maquet, B. Carré, F. Lepetit, J.-F. Hergott, T. Ruchon, L. Argenti, R. Taïeb, F. Martín, and P. Salières, Science (80-. ). **354**, 734 (2016).

[23] A. Kaldun, A. Blättermann, V. Stooß, S. Donsa, H. Wei, R. Pazourek, S. Nagele, C. Ott, C. D. Lin, J. Burgdörfer, and T. Pfeifer, Science (80-. ). **354**, 738 (2016).

[24] M. Ossiander, F. Siegrist, V. Shirvanyan, R. Pazourek, A. Sommer, T. Latka, A. Guggenmos, S. Nagele, J. Feist, J. Burgdörfer, R. Kienberger, and M. Schultze, Nat. Phys. **13**, 280 (2017).

[25] A. N. Artemyev, L. S. Cederbaum, and P. V Demekhin, Phys. Rev. A **96**, 033410 (2017).

[26] G. Mouloudakis and P. Lambropoulos, J. Phys. B At. Mol. Opt. Phys. **51**, 01LT01 (2018).

[27] G. Mouloudakis and P. Lambropoulos, Eur. Phys. J. D **72**, 226 (2018).

[28] M. Nagasono, E. Suljoti, A. Pietzsch, F. Hennies, M. Wellhöfer, J.-T. Hoeft, M. Martins, W. Wurth, R. Treusch, J. Feldhaus, J. R. Schneider, and A. Föhlisch, Phys. Rev. A **75**, 051406 (2007).

[29] L. Young, E. P. Kanter, B. Krässig, Y. Li, A. M. March, S. T. Pratt, R. Santra, S. H. Southworth, N. Rohringer, L. F. DiMauro, G. Doumy, C. A. Roedig, N. Berrah, L. Fang, M. Hoener, P. H. Bucksbaum, J. P. Cryan, S. Ghimire, J. M. Glownia, D. A. Reis, J. D. Bozek, C. Bostedt, and M. Messerschmidt, Nature **466**, 56 (2010).

[30] A. Hishikawa, M. Fushitani, Y. Hikosaka, A. Matsuda, C.-N. Liu, T. Morishita, E. Shigemasa, M. Nagasono, K. Tono, T. Togashi, H. Ohashi, H. Kimura, Y. Senba, M. Yabashi, and T. Ishikawa, Phys. Rev. Lett. **107**, 243003 (2011).

[31] E. P. Kanter, B. Krässig, Y. Li, A. M. March, P. Ho, N. Rohringer, R. Santra, S. H.

Southworth, L. F. DiMauro, G. Doumy, C. A. Roedig, N. Berrah, L. Fang, M. Hoener, P. H. Bucksbaum, S. Ghimire, D. A. Reis, J. D. Bozek, C. Bostedt, M. Messerschmidt, and L. Young, Phys. Rev. Lett. **107**, 233001 (2011).

[32]  C. Weninger, M. Purvis, D. Ryan, R. A. London, J. D. Bozek, C. Bostedt, A. Graf, G. Brown, J. J. Rocca, and N. Rohringer, Phys. Rev. Lett. **111**, 233902 (2013).

[33]  M. Žitnik, A. Mihelič, K. Bučar, M. Kavčič, J.-E. Rubensson, M. Svanquist, J. Söderström, R. Feifel, C. Såthe, Y. Ovcharenko, V. Lyamayev, T. Mazza, M. Meyer, M. Simon, L. Journel, J. Lüning, O. Plekan, M. Coreno, M. Devetta, M. Di Fraia, P. Finetti, R. Richter, C. Grazioli, K. C. Prince, and C. Callegari, Phys. Rev. Lett. **113**, 193201 (2014).

[34]  K. C. Prince, E. Allaria, C. Callegari, R. Cucini, G. De Ninno, S. Di Mitri, B. Diviacco, E. Ferrari, P. Finetti, D. Gauthier, L. Giannessi, N. Mahne, G. Penco, O. Plekan, L. Raimondi, P. Rebernik, E. Roussel, C. Svetina, M. Trovò, M. Zangrando, M. Negro, P. Carpeggiani, M. Reduzzi, G. Sansone, A. N. Grum-Grzhimailo, E. V Gryzlova, S. I. Strakhova, K. Bartschat, N. Douguet, J. Venzke, D. Iablonskyi, Y. Kumagai, T. Takanashi, K. Ueda, A. Fischer, M. Coreno, F. Stienkemeier, Y. Ovcharenko, T. Mazza, and M. Meyer, Nat. Photonics **10**, 176 (2016).

[35]  J. R. Harries, H. Iwayama, S. Kuma, M. Iizawa, N. Suzuki, Y. Azuma, I. Inoue, S. Owada, T. Togashi, K. Tono, M. Yabashi, and E. Shigemasa, Phys. Rev. Lett. **121**, 263201 (2018).

[36]  M. Domke, K. Schulz, G. Remmers, G. Kaindl, and D. Wintgen, Phys. Rev. A **53**, 1424 (1996).

[37]  J. M. Rost, K. Schulz, M. Domke, and G. Kaindl, J. Phys. B At. Mol. Opt. Phys. **30**, 4663 (1997).

[38]  L. B. Madsen, P. Schlagheck, and P. Lambropoulos, Phys. Rev. A **62**, 062719 (2000).

[39]  M. B. Gaarde, C. Buth, J. L. Tate, and K. J. Schafer, Phys. Rev. A **83**, 013419 (2011).

[40]  K. Tiedtke, A. Azima, N. von Bargen, L. Bittner, S. Bonfigt, S. Düsterer, B. Faatz, U.

Frühling, M. Gensch, C. Gerth, N. Guerassimova, U. Hahn, T. Hans, M. Hesse, K. Honkavaar, U. Jastrow, P. Juranic, S. Kapitzki, B. Keitel, T. Kracht, M. Kuhlmann, W. B. Li, M. Martins, T. Núñez, E. Plönjes, H. Redlin, E. L. Saldin, E. A. Schneidmiller, J. R. Schneider, S. Schreiber, N. Stojanovic, F. Tavella, S. Toleikis, R. Treusch, H. Weigelt, M. Wellhöfer, H. Wabnitz, M. V Yurkov, and J. Feldhaus, New J. Phys. **11**, 023029 (2009).

[41]  K. Tiedtke, J. Feldhaus, U. Hahn, U. Jastrow, T. Nunez, T. Tschentscher, S. V Bobashev, A. A. Sorokin, J. B. Hastings, S. Möller, L. Cibik, A. Gottwald, A. Hoehl, U. Kroth, M. Krumrey, H. Schöppe, G. Ulm, and M. Richter, J. Appl. Phys. **103**, 094511 (2008).

[42]  S. Düsterer, M. Rehders, A. Al-Shemmary, C. Behrens, G. Brenner, O. Brovko, M. DellAngela, M. Drescher, B. Faatz, J. Feldhaus, U. Frühling, N. Gerasimova, N. Gerken, C. Gerth, T. Golz, A. Grebentsov, E. Hass, K. Honkavaara, V. Kocharian, M. Kurka, T. Limberg, R. Mitzner, R. Moshammer, E. Plönjes, M. Richter, J. Rönsch-Schulenburg, A. Rudenko, H. Schlarb, B. Schmidt, A. Senftleben, E. A. Schneidmiller, B. Siemer, F. Sorgenfrei, A. A. Sorokin, N. Stojanovic, K. Tiedtke, R. Treusch, M. Vogt, M. Wieland, W. Wurth, S. Wesch, M. Yan, M. V. Yurkov, H. Zacharias, and S. Schreiber, Phys. Rev. Spec. Top. - Accel. Beams **17**, 120702 (2014).

[43]  T. Pfeifer, Y. Jiang, S. Düsterer, R. Moshammer, and J. Ullrich, Opt. Lett. **35**, 3441 (2010).

[44]  S. Roling, B. Siemer, M. Wöstmann, H. Zacharias, R. Mitzner, A. Singer, K. Tiedtke, and I. A. Vartanyants, Phys. Rev. Spec. Top. - Accel. Beams **14**, 080701 (2011).

[45]  G. Brenner, S. Kapitzki, M. Kuhlmann, E. Ploenjes, T. Noll, F. Siewert, R. Treusch, K. Tiedtke, R. Reininger, M. D. Roper, M. A. Bowler, F. M. Quinn, and J. Feldhaus, Nucl. Instruments Methods Phys. Res. Sect. A Accel. Spectrometers, Detect. Assoc. Equip. **635**, S99 (2011).

[46]  J. A. R. Samson, Z. X. He, L. Yin, and G. N. Haddad, J. Phys. B At. Mol. Opt. Phys. **27**, 887 (1994).

[47]   K. Meyer, C. Ott, P. Raith, A. Kaldun, Y. Jiang, A. Senftleben, M. Kurka, R. Moshammer, J. Ullrich, and T. Pfeifer, Phys. Rev. Lett. **108**, 098302 (2012).

[48]   W. F. Schlotter, F. Sorgenfrei, T. Beeck, M. Beye, S. Gieschen, H. Meyer, M. Nagasono, A. Föhlisch, and W. Wurth, Opt. Lett. **35**, 372 (2010).

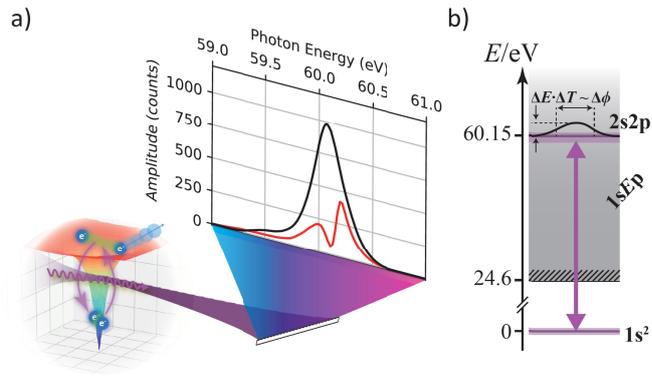

Fig. 1: Conceptual design of the measurement and the physics scheme. **a)** An intense XUV pulse (violet shading and arrows) is focused through a dense gas medium, tuned to the resonance of a single-photon two-electron excitation. The transmitted XUV light is spectrally dispersed after a grating (cyan to magenta color gradient) and measured with an XUV-CCD camera. The average transmitted experimental spectra without (black curve) and with (red curve) the helium target are plotted. **b)** Energy level scheme of the helium atom. The two-electron ground state ($1s^2$) is resonantly coupled (violet arrow and shading) to the two-electron excited state ($2s2p$), which is embedded in the single-electron ionization continuum ($1sEp$). Strong-field dressing leads to a transient energy shift $\Delta E$ during the interaction with the pulse duration $\Delta T$ which translates into a phase shift $\Delta \phi$ of the dipole response.

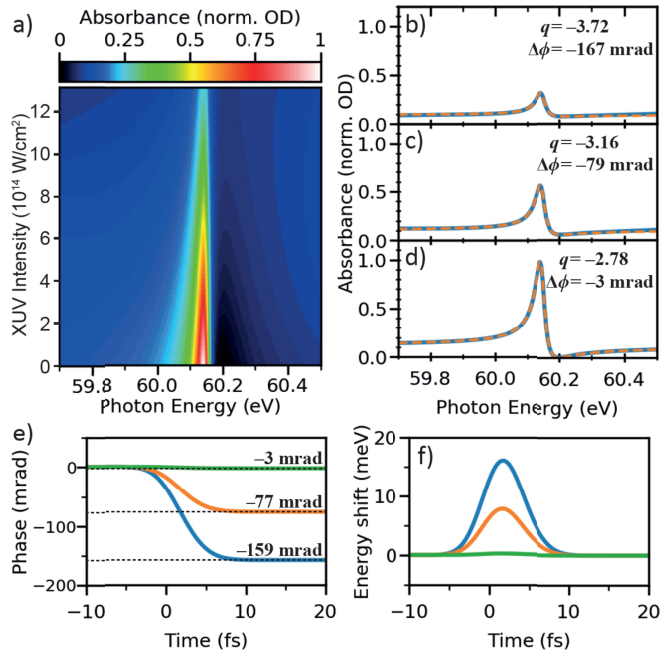

Fig. 2: Model simulation of transient excitation and dressing of the 2s2p Fano line in helium with a 5-fs-duration (FWHM) Gaussian-shaped pulse centered at photon energy 60.10 eV. **a)** Intensity dependence of the resonant absorption, indicated in color scale in units of normalized optical density (OD). **b)-d)** Lineouts (blue solid lines) of the resonant absorption [as shown in a)] for a high [b] $1.1 \cdot 10^{15}$ W/cm², intermediate [c] $5.6 \cdot 10^{14}$ W/cm², and low [d] $1.9 \cdot 10^{13}$ W/cm² XUV intensity. The fit of the Fano line shape (orange dashed lines) reveals a change of $q$ and corresponding relative phase shift $\Delta\phi$ due to the dressing during the pulse. **e)** Time-dependent phase evolution $\phi_e(t)$ of the 2s2p excited state for low ($1.9 \cdot 10^{13}$ W/cm²; green line), intermediate ($5.6 \cdot 10^{14}$ W/cm²; orange line), and high ($1.1 \cdot 10^{15}$ W/cm²; blue line) XUV intensity. The numerical values and horizontal dotted lines denote the phase shifts after the dressing, which agree well with the results obtained through fitting the line profiles for the respective photon fluence [b)–d)]. **f)** Transient energy shift $\Delta E(t) = -\hbar \cdot \partial/\partial t [\phi_e(t)]$ for the three XUV intensity settings with the line colors corresponding to the curves shown in e).

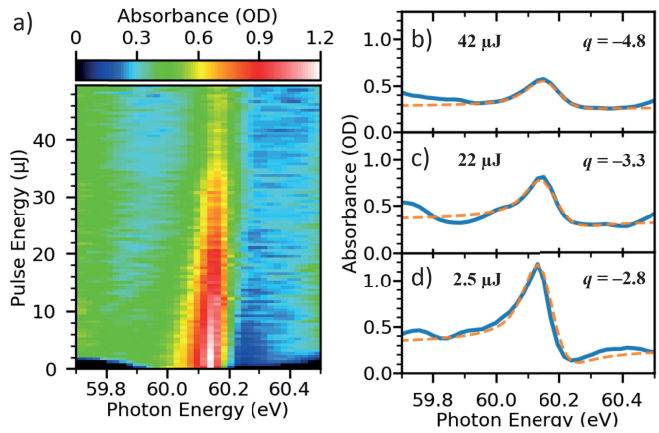

Fig. 3: Measured absorbance of intense XUV FEL pulses transmitted through a dense helium target, retrieved from an average over at least 50 and up to 300 single-shot FEL spectra for each pulse-energy bin. **a)** Pulse-energy dependence of the resonant absorption, indicated in color scale in units of the optical density (OD). **b)-d)** Lineouts (blue solid line) of the resonant absorption [as shown in (a)] for a high [(b) 42 µJ], intermediate [(c) 22 µJ], and low [(d) 2.5 µJ] pulse energy. The fit of a Fano spectral profile (orange dashed line) quantifies the increasing magnitude of $q$ with increasing pulse energy, which agrees well with the predicted results of the model calculation in Fig. 2. In the weak-field limit [(d)] the orange dashed line is drawn according to the resonance parameters reported in literature ($q = -2.8$).